\documentclass[aps,prc,twocolumn,superscriptaddress,showpacs,floatfix,nofootinbib]{revtex4-1}
\usepackage{amsmath,graphicx,color}

\begin{document}
\title{Accessing the speed of sound in relativistic ultracentral nucleus-nucleus collisions using the mean transverse momentum}
\author{Fernando G. Gardim}
\affiliation{Instituto de Ci\^encia e Tecnologia, Universidade Federal de Alfenas, 37715-400 Po\c cos de Caldas, MG, Brazil}
\affiliation{ Illinois Center for Advanced Studies of the Universe, Department of Physics, University of Illinois at
Urbana-Champaign, Urbana, Illinois 61801, USA}
\author{Andre V. Giannini}
\affiliation{Faculdade de Ci\^encias Exatas e Tecnologia, Universidade Federal da Grande Dourados, 79804-970 Dourados, MS, Brazil}
\affiliation{Departamento de F\'isica, Universidade do Estado de Santa Catarina, 89219-710 Joinville, SC, Brazil}
\author{Jean-Yves Ollitrault}
\affiliation{Universit\'e Paris Saclay, CNRS, CEA, Institut de physique th\'eorique, 91191 Gif-sur-Yvette, France}
\begin{abstract}
It has been argued that the speed of sound of the strong interaction at high temperature can be measured using the  variation of the mean transverse momentum with the particle multiplicity in ultracentral heavy-ion collisions. 
We test this correspondence by running hydrodynamic simulations at zero impact parameter with several equations of state, at several  colliding energies from 0.2~TeV to 15~TeV per nucleon pair.  
The correspondence is found to be precise and robust for a smooth, boost-invariant fluid and an ideal detector. 
We discuss the differences between this ideal setup and an actual experiment. 
We conclude that the extraction of the speed of sound from data is reliable, and that the main uncertainty comes from our poor knowledge of the distribution of density fluctuations at the early stages of the collision. 
\end{abstract}
\maketitle

\section{Introduction}
One of the primary goals of ultrarelativistic nucleus-nucleus collisions is to infer the thermodynamic properties of the quark-gluon plasma which they create.  
This is a fundamental question, as it addresses the thermodynamics of the strong interaction~\cite{Ratti:2018ksb}. 
The general methodology is clear. 
It is established that the evolution of the system produced in these collisions is described by relativistic viscous hydrodynamics~\cite{Gale:2013da,Romatschke:2017ejr}. 
Now, thermodynamic properties (equation of state and transport coefficients) are an input of hydrodynamic equations. 
By comparing the particle distributions measured in experiment with those calculated in hydrodynamic models, one may therefore constrain these properties~\cite{Bernhard:2016tnd,Pang:2016vdc,Gardim:2019xjs,Nijs:2020roc,JETSCAPE:2020mzn}. 

In a hydrodynamic simulation, the temperature depends on time and space coordinates. 
Specifically, it decreases as a function of time because the system expands into the vacuum. 
The temperature at a given early time decreases as one moves from the center toward the edges of the fireball. 
The hydrodynamic expansion therefore spans a wide range of temperatures, and one does not expect the correspondence between experimental data and thermodynamic properties to be simple. 
This has motivated the development of global comparisons between theory and experiment, in which the thermodynamic properties are constrained from collider data through Bayesian analyses~\cite{Bernhard:2016tnd,Nijs:2020roc,JETSCAPE:2020mzn}. 

As for the equation of state, however, a simple proportionality relation has been observed~\cite{Gardim:2019xjs} 
in hydrodynamic simulations  
between the mean transverse momentum of outgoing particles, $\langle p_T\rangle$, and an effective temperature $T_{\rm eff}$, confirming an old idea put forward by Van Hove~\cite{VanHove:1982vk}:
\begin{equation}
\label{ptteff}
\langle p_T\rangle\simeq 3\, T_{\rm eff},
\end{equation}
%While it is natural to expect this relation in the limit of an ultrarelativistic massless ideal gas, we will instead obtain this relation through phenomenological study through hydrodynamic simulations.
where the proportionality factor $3$ is purely phenomenological, while the effective temperature has a precise definition, which will be recalled below. 
$T_{\rm eff}$ always lies between the final and initial temperatures, and roughly corresponds to the average temperature at a time of the order of the nuclear radius~\cite{Gardim:2019xjs}. 

This proportionality relation becomes particularly interesting when applied to ultracentral collisions~\cite{Gardim:2019brr}. 
Ultracentral collisions are defined as those producing the largest number of particles~\cite{Luzum:2012wu,CMS:2013bza}. 
The impact parameter of an ultracentral collision is close to zero, but the multiplicity can vary by up to 10-15\% due to quantum fluctuations, thus an increase of $\langle p_T \rangle$ is expected as the multiplicity increases, if the matter exhibits fluid-like behavior. This increase is a genuine  hydrodynamic effect~\cite{Gardim:2019brr}.
Ultracentral collisions are therefore a laboratory to investigate the effect of a change in the system density, at fixed geometry~\cite{Samanta:2023amp}. 
The temperature increase induced by an increase in density is determined by the speed of sound $c_s$, defined by $c_s^2=d\ln T/d\ln s$, where $s$ is the entropy density~\cite{Ollitrault:2007du}. 
Since the entropy density is proportional to the particle multiplicity $N_{ch}$, one readily obtains, using Eq.~(\ref{ptteff}):
\begin{equation}
\label{csdata}
c_s^2(T_{\rm eff})=\frac{d\ln\langle p_T\rangle }{d\ln N_{ch}}. 
\end{equation} 
The CMS collaboration has recently measured the speed of sound at $T_{\rm eff}=219\pm 8$~MeV with this method, and the result 
$c_s^2=0.241\pm 0.016$ is in perfect agreement with lattice QCD calculations~\cite{CMS:2024sgx}.

The primary goal of this paper is to assess more precisely the validity of Eqs.~(\ref{ptteff}) and (\ref{csdata}), which form the theoretical basis of the CMS analysis, by means of systematic hydrodynamic calculations.
This reassessment is necessary in the light of recent state-of-the-art hydrodynamic calculations~\cite{Nijs:2023bzv} suggesting that these equations are not precise. 

For the sake of simplicity, we simulate, as a first step, an ideal situation consisting of Pb+Pb collisions at zero impact parameter, $b=0$, with a smooth initial density profile, and assuming exact longitudinal boost invariance~\cite{Bjorken:1982qr}. 
Within this simplified setup, we check the robustness of Eqs.~(\ref{ptteff}) and (\ref{csdata}) against variations of the equation of state and collision energy.
We then discuss the differences between our ideal setup and an actual experiment and their impact on the determination of $c_s$.

\section{Method}
\label{s:method}

\begin{figure}[h]
\begin{center}
\includegraphics[width=\linewidth]{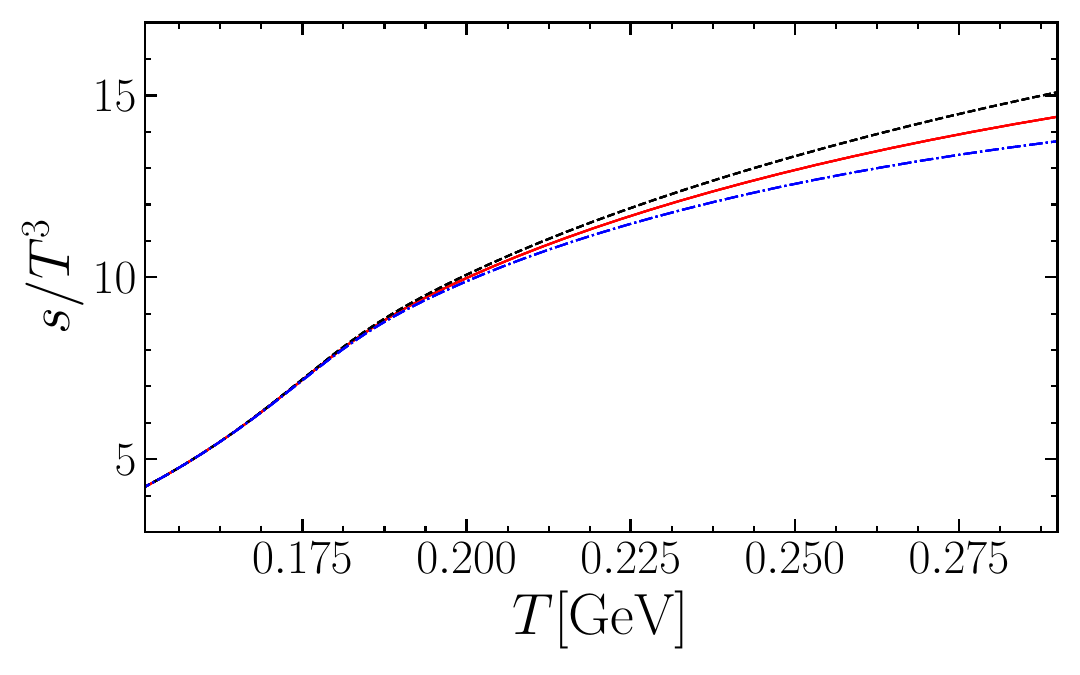} 
\end{center}
\caption{
  Representation of the three equations of state implemented in our calculations. $s$ denotes the entropy density. 
The central curve is the first-principles calculation by the hotQCD collaboration~\cite{HotQCD:2014kol}.
  The upper and lower curves correspond to the equations of state obtained after applying the modification (\ref{modifeos}) with $\alpha=\pm 1.54$.
The entropy density is related to the pressure through the thermodynamic identity $s=dP/dT$. 
}
\label{fig:eos}
\end{figure} 

The default equation of state of our hydrodynamic calculation is that obtained through lattice calculations by the hotQCD collaboration~\cite{HotQCD:2014kol}.
It matches precisely that of a hadronic resonance gas at low temperatures, which is crucial in order to ensure energy and momentum conservation when the fluid is converted into hadrons. 

We choose to study only smooth, small variations around this equation of state. 
The reason why we consider smooth variations is twofold: 
First, our approach involves an effective temperature, which is averaged over the whole system, as we shall see shortly. 
Because of this averaging, we do not expect it to catch sharp variations of the equation of state around the effective temperature. 
We rather expect results to becomes less precise as the rate of variation of thermodynamic quantities increases. 
We will see a hint of this effect in our results presented in Sec.~\ref{s:results}. 
The second reason is that the hydrodynamic description itself breaks down in the presence of large gradients~\cite{Baier:2007ix}. 
We have noticed in earlier works that even sharp variations of the transport coefficients (which enter the hydrodynamic equations only as small corrections~\cite{Baier:2007ix}) generate unstable results~\cite{Gardim:2020mmy}. 
The reason why we only consider small variations around the lattice equation of state is that very different equations of state (which were implemented in Ref.~\cite{Gardim:2019xjs} in order to test the robustness of the effective hydrodynamic description) are already ruled out by data.
Our priority here is to test whether we are able to capture small changes in the speed of sound through experimental data. 

A further requirement on the equation of state is that it is not  modified near the freeze-out temperature $T=T_F$ at which the fluid is converted into hadrons. 
We choose the following modification of the pressure, which meets all the previous requirements: 
\begin{equation}
\label{modifeos}
  P(T)\rightarrow P(T)+\alpha(T-T_F)^4, 
\end{equation}
where $T_F=151$~MeV~\cite{Moreland:2018gsh} is the freeze-out temperature in our hydrodynamic calculation, so that we use Eq.~(\ref{modifeos}) only for $T>T_F$, and $\alpha$ is a dimensionless constant.
Eq.~(\ref{modifeos}) ensures that the modified entropy density and speed of sound vary smoothly with the temperature, and are not modified at $T=T_F$.
In the limit $T\to\infty$, the ratio $P(T)/T^4$  is proportional to the number of degrees of freedom of quarks and gluons, so that $\alpha$ physically represents a change in the number of degrees of freedom~\cite{Nijs:2023bzv}. 
We implement two opposite values $\alpha= 1.54$ and $\alpha=-1.54$, in addition to the default value $\alpha=0$. 
As will be shown below, a positive $\alpha$ implies a softer equation of state, with a lower speed of sound, while a negative $\alpha$ implies a harder equation of state. 
The three equations of state implemented in our calculations are illustrated in Fig.~\ref{fig:eos}, where we choose to plot the entropy density rather than the pressure or energy density, since the entropy density is more directly accessible from experimental data~\cite{Gardim:2019xjs}. 

The setup of our hydrodynamic calculation is that of the Duke analysis (Table IV of Ref.~\cite{Moreland:2018gsh}), which is calibrated to reproduce LHC data on p+Pb and Pb+Pb collisions at $\sqrt{s}_{NN}=5.02$~TeV. 

The only difference lies in the initial condition. 
Instead of generating a large number of events with randomly fluctuating initial density profiles, we run a single event with a smooth initial density. 
This smooth density is obtained by generating 1000 events at $b=0$ using the T$_{\rm R}$ENTo model~\cite{Moreland:2014oya} (tuned as in Table IV of Ref.~\cite{Moreland:2018gsh}) and averaging the energy density over all events~\cite{Song:2010mg}. 
The interest of working with a smooth density profile is that we can assess precisely the effect of a variation in multiplicity,  by rescaling the initial energy density profile by a multiplicative constant. 
Such a variation in multiplicity can arise either from  fluctuations in ultracentral collisions at fixed collision energy, which are the focus of this work, or from a variation of the collision energy. 
Our setup allows us to study both effects simultaneously. 

This initial condition is then evolved through the hydrodynamic code MUSIC~\cite{Schenke:2010nt,Schenke:2010rr,Huovinen:2012is,Paquet:2015lta} after a free-streaming stage. 
The parameters of the hydrodynamic calculation (time of the free-streaming evolution, shear and bulk viscosity) are again taken from Table IV of Ref.~\cite{Moreland:2018gsh}. 

When the fluid cools down to the temperature $T_F$, it is locally converted~\cite{Petersen:2008dd} into a hadron gas, and interactions in the subsequent hadronic phase are modeled with the UrQMD transport code~\cite{Bass:1998ca,Bleicher:1999xi}. 
The sampling of the fluid into particles is a random discretization process which destroys the smoothness of the fluid description. 
We restore the smoothness by repeating the sampling and running the transport calculation 1000 times, which reduces the statistical errors~\cite{Gardim:2011qn}. 
%who did this oversampling first? I doubt that we invented this. 

For each hydrodynamic simulation, we evaluate the total energy $E$ and entropy $S$ of the fluid  (more precisely, the quantities per unit rapidity, $dE/dy$ and $dS/dy$, at $y=0$) at the temperature where it is converted into hadrons, by integrating over the hypersurface $T=T_F$ as explained in Ref.~\cite{Gardim:2019xjs}. 
Note that $E$ denotes the energy in the laboratory frame, not in the fluid rest frame. It contains the kinetic energy associated with transverse collective flow. 
The effective temperature $T_{\rm eff}$ and the effective volume per unit rapidity $V_{\rm eff}$ (which should be denoted by $dV_{\rm eff}/dy$, but we keep the original notation for the sake of simplicity) are obtained by solving the coupled system of equations: 
\begin{eqnarray}
  \label{effective}
E&=&\epsilon(T_{\rm eff}) V_{\rm eff},\cr
S&=&s(T_{\rm eff}) V_{\rm eff}, 
\end{eqnarray}
where the functions $\epsilon(T)$ and $s(T)$ correspond to the equation of state used in the hydrodynamic calculation. 
In order to understand the physical meaning of $T_{\rm eff}$ and $V_{\rm eff}$, consider first the following simplified situation: 
1) The fluid is initially at rest and homogeneous, so that the initial temperature is uniform. 2) Dissipative effects are small, so that $S$ is conserved. 3) There is no longitudinal expansion, so that $E$ is also conserved. 
Then the values of $E$ and $S$ evaluated at the end of the hydrodynamic calculation coincide with the initial values.  
This in turn implies that $T_{\rm eff}$ and $V_{\rm eff}$ coincide with the initial temperature and the initial volume. 

The largest correction to this simplified picture arises from the longitudinal expansion~\cite{Bjorken:1982qr}, which results in a value of $T_{\rm eff}$ significantly smaller than the initial temperature (and a value of $V_{\rm eff}$  larger than the initial volume). 
Note, however, that $T_{\rm eff}$ is always larger than the freeze-out temperature, because $E$ contains the kinetic energy associated with transverse flow. 
%$T_{\rm eff}$ and $V_{\rm eff}$ defined by Eq.~(\ref{effective}) are the temperature and volume of a fluid at rest that would have the same energy and entropy as at the end of the hydrodynamic evolution. 
%From these definitions, one sees that $T_{\rm eff}$ coincides with $T_F$ only if the fluid is at rest, and transverse flow always implies $T_{\rm eff}>T_F$. 
%In the case of a uniform initial temperature $T_i$, one always has  $T_{\rm eff}<T_i$ because the energy of the fluid decreases throughout the hydrodynamic phase due to longitudinal cooling~\cite{Bjorken:1982qr}. 

We then evaluate the charged multiplicity density per unit pseudorapidity $dN_{ch}/d\eta$ and the average transverse momentum $\langle p_T\rangle$ of charged particles at the end of the evolution, that is, after the last interactions in the hadronic phase and after decays of hadronic resonances.
To the extent that the pseudorapidity coincides with the rapidity, and that the hadron multiplicity after the hadronic phase is proportional to that at $T_F$,  $dN_{ch}/d\eta$ is proportional to the final fluid entropy $S$. 
If, in addition, the effective volume $V_{\rm eff}$ is constant, and if the entropy per charged particle at freeze-out is a constant~\cite{Hanus:2019fnc}, then $dN_{ch}/d\eta$ is proportional to $s(T_{\rm eff})$. 
If $\langle p_T\rangle/T_{\rm eff}$ is a universal constant, as observed in Ref.~\cite{Gardim:2019xjs}, and in a hybrid framework for small systems~\cite{Gardim:2022yds}, then Eq.~(\ref{csdata}) automatically follows. 
Thus, the first step is to check whether these two properties are satisfied for all equations of state and colliding energies. 

\section{Results}
\label{s:results}

\begin{figure}[h]
\begin{center}
\includegraphics[width=.95\linewidth]{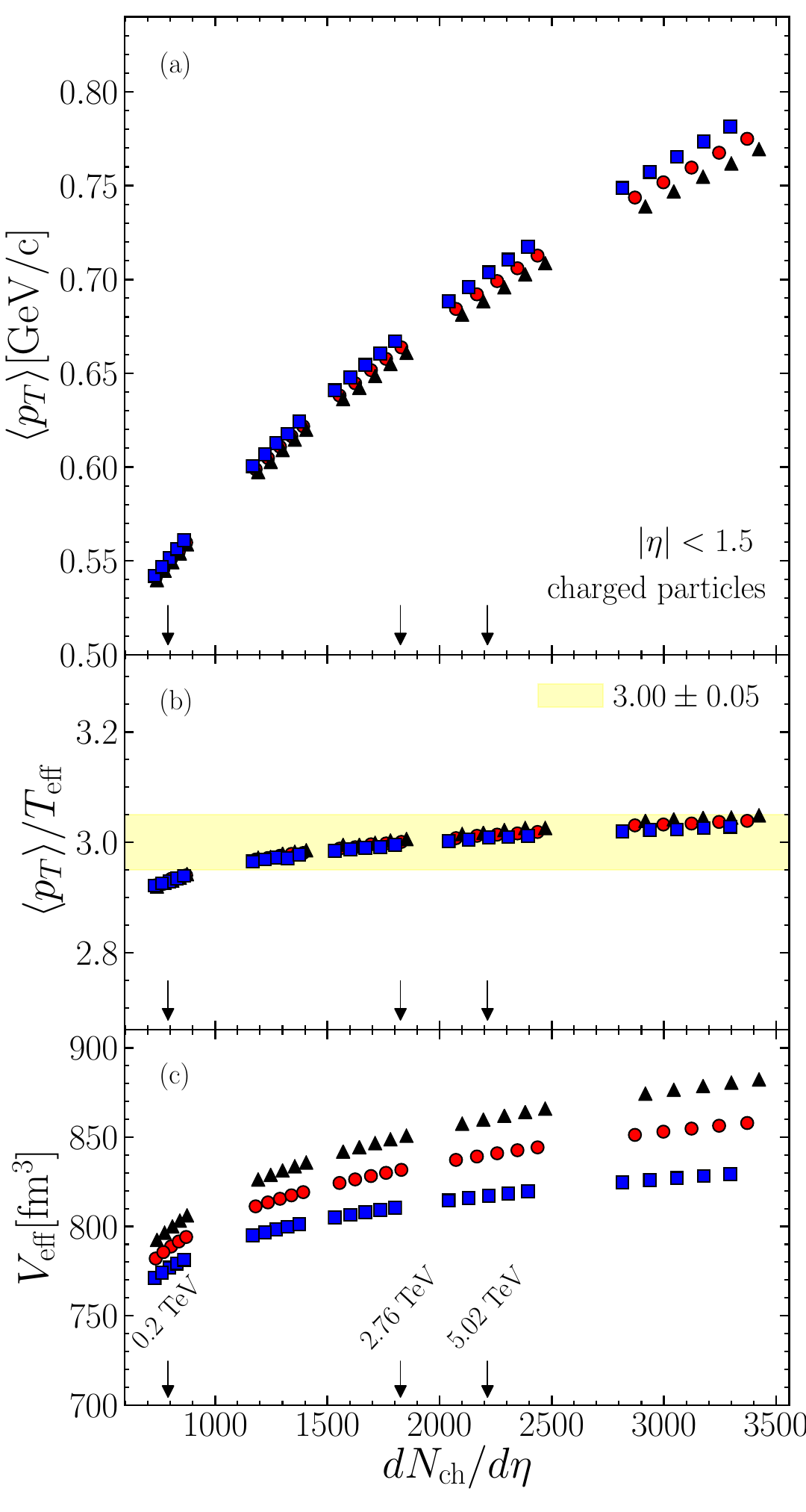} 
\end{center}
\caption{
Results of the hydrodynamic simulations. 
The three different sets of symbols correspond to the three equations of state displayed in Fig.~\ref{fig:eos}. 
Results are plotted as a function of the charged multiplicity density in the pseudorapidity interval $|\eta|<1.5$. 
(a) Average transverse momentum of charged hadrons. 
(b) Ratio of average transverse momentum to effective temperature.
The yellow band, which encompasses all the points, is our estimate of the error bar. 
(c) Effective volume per unit rapidity. 
}
\label{fig:validation}
\end{figure} 

Results are displayed in Fig.~\ref{fig:validation}. 
We select particles in a broad pseudorapidity window $|\eta|<1.5$, for reasons that will be detailed below. 
For each of the three equations of state, the simulation has been run for 25 values of $dN_{ch}/d\eta$, grouped in 5 sets of 5 points. 
Each set of 5 points is meant to simulate multiplicity fluctuations at fixed collision energy, and will be used to infer the speed of sound using Eq.~(\ref{csdata}) by a linear fit. 
The 5 sets correspond to different collision energies. 
They span approximately the range $0.2<\sqrt{s_{NN}}<15$~TeV, from the top RHIC energy up to three times the current LHC energy. 

The values of $dN_{ch}/d\eta$ corresponding to specific collision energies are indicated by arrows in Fig.~\ref{fig:validation}.
These values are obtained by extrapolating the measured multiplicities at $\sqrt{s_{NN}}=0.2$~TeV~\cite{PHOBOS:2002lqa}, 
$2.76$~TeV~\cite{ALICE:2010mlf} and $5.02$~TeV~\cite{ALICE:2015juo} in the following way: 
First, we extrapolate linearly, using the first two centrality bins, down to 0\% centrality, corresponding approximately to collisions at b=0. 
Next, we take into account the different $\eta$ intervals ($|\eta|<1$ for Ref.~\cite{PHOBOS:2002lqa}, $|\eta|<0.5$ for Refs.~\cite{ALICE:2010mlf,ALICE:2015juo}, $|\eta|<1.5$ for our calculation) by rescaling the multiplicity by the same factor as in our hydrodynamic calculation (it is less than a 5\% increase in practice). 
Then, we extrapolate results from Au+Au collisions~\cite{PHOBOS:2002lqa} to Pb+Pb collisions assuming that $dN_{ch}/d\eta$ is proportional to the mass number of colliding nuclei, that is, by a factor $208/197$. 
Finally, the values of $dN_{ch}/d\eta$ can be extrapolated beyond the current LHC energy using $dN_{ch}/d\eta\propto s_{NN}^{0.155}$~\cite{ALICE:2015juo}.

For a given equation of state, $\langle p_T\rangle$ increases as a function of $dN_{ch}/d\eta$ (Fig.~\ref{fig:validation} (a)). 
This increase is driven by an increase in the effective temperature $T_{\rm eff}$, and the ratio  $\langle p_T\rangle/T_{\rm eff}$ is remarkably constant (Fig.~\ref{fig:validation} (b)), as already found in Refs.~\cite{Gardim:2019xjs,Gardim:2022yds}. 
It does not depend on the equation of state, and increases very  mildly as a function of colliding energy. 
Our results on the speed of sound, which will be shown below, are obtained by assuming $\langle p_T\rangle/T_{\rm eff}=3.00\pm 0.05$,  consistent with the choice made by CMS~\cite{CMS:2024sgx}, except for the first set of points, corresponding to the top RHIC energy, for which we assume a slightly smaller value $\langle p_T\rangle/T_{\rm eff}=2.93\pm 0.05$.  

Unlike the ratio $\langle p_T\rangle/T_{\rm eff}$, the effective volume  $V_{\rm eff}$ depends somewhat on the equation of state (Fig.~\ref{fig:validation} (c)). 
 A harder equation of state results in a smaller effective volume, which can be attributed to a more rapid expansion~\cite{Gardim:2019xjs}. 
 For a given equation of state, the effective volume increases somewhat with the particle multiplicity, but this is a very mild increase.

Note that we do not vary the transport coefficients and initial conditions. 
This study was carried out in detail in Ref.~\cite{Gardim:2019xjs}, where the ratio $\langle p_T\rangle/T_{\rm eff}$ was found to be insensitive to shear and bulk viscosity, and to collision centrality (different collision centralities correspond in practice to different initial conditions). 
Therefore, even though the initial density profile in our calculation was inferred from LHC data~\cite{Moreland:2018gsh}, the value of $\langle p_T\rangle/T_{\rm eff}$ is robust with respect to the choice made, and our results should also apply at other colliding energies. 

\begin{figure}[h]
\begin{center}
\includegraphics[width=\linewidth]{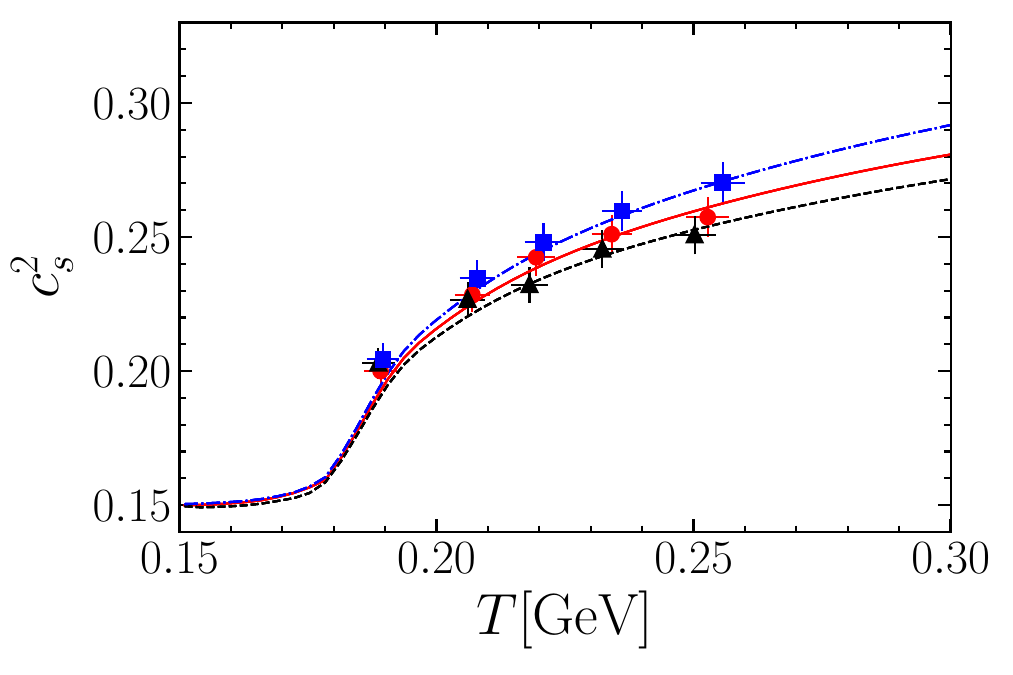} 
\end{center}
\caption{
Lines: speed of sound squared as a function of temperature for the three equations of state displayed in Fig.~\ref{fig:eos}, calculated using $c_s^2=(d\ln s/d\ln T)^{-1}$. 
Note that the ordering of the curves is inverted compared to Fig.~\ref{fig:eos}: 
The hardest equation of state, with the largest speed of sound, is that with the lowest entropy density. 
Symbols: values of $T$ and $c_s^2(T)$ inferred from Eqs.~(\ref{ptteff}) and (\ref{csdata}).
Vertical bars are statistical errors estimated with the jackknife method.
Horizontal bars correspond to the $\pm 0.05$ uncertainty on $\langle p_T\rangle /T_{\rm eff}$. 
}
\label{fig:cs2}
\end{figure} 

%Thus the conditions for extracting the speed of sound from data, that both $\langle p_T\rangle/T_{\rm eff}$ and $V_{\rm eff}$ are independent of multiplicity, are met to a good approximation. 
We finally compute $c_s^2$ using  Eq.~(\ref{csdata}), where the derivative is evaluated by fitting each set of 5 points in Fig.~\ref{fig:validation} (a)  using $\ln\langle p_T\rangle=c_s^2\ln N_{ch}+{\rm constant}$. 
The corresponding effective temperature is then evaluated from $\langle p_T\rangle$ using the proportionality relations above. 

Our main result is displayed in Fig.~\ref{fig:cs2}. 
The values of $c_s^2(T)$ extracted from Eq.~(\ref{csdata}) agree with the corresponding equations of state, within statistical error bars, for a broad range of temperatures. 
The value of $c_s^2$ is slightly overestimated, by $\sim 0.01$, only for the lowest temperature, corresponding to the top RHIC energy. 
This slight discrepancy can likely be assigned to the sharp variation of $c_s^2$ around this temperature. 
As anticipated in Sec.~\ref{s:method}, the averaging implied by the effective temperature becomes less effective in this case. 

Note that if $\langle p_T\rangle/T_{\rm eff}$ and $V_{\rm eff}$ were strictly independent of multiplicity in Fig.~\ref{fig:validation}, the agreement between the symbols and the curves in Fig.~\ref{fig:cs2} would be trivial, as it simply follows from the thermodynamic definition of $c_s$. 
But both quantities increase slightly with multiplicity, which implies that $\langle p_T\rangle$ increases somewhat faster than $T_{\rm eff}$, and that $dN_{ch}/d\eta$ increases somewhat faster than the entropy density $s(T_{\rm eff})$. 
Interestingly, these effects compensate one another to some extent, and the determination of the speed of sound is even more precise than one would have expected. 

\section{Discussion}
\label{s:discussion}

We now discuss the differences between our idealized setup and an actual experiment.

\subsection{Centrality determination}

First, our calculation is carried out at zero impact parameter, $b=0$.
But $b$ is not measured, and one cannot select events with $b=0$ (or smaller than some number) in practice.  
A specific observable is used as a centrality classifier, for instance the transverse energy in a forward calorimeter~\cite{CMS:2024sgx}. 
In each centrality bin, one then measures the average multiplicity and transverse momentum. 
The limit $b=0$ is only approached asymptotically for the largest multiplicities. 
An interpolation formula, which describes how this limit is approached, has been derived in~\cite{Gardim:2019brr} using the general methodology introduced in \cite{Das:2017ned}. 
It is implemented in the CMS analysis~\cite{CMS:2024sgx}.

Nijs and van der Schee have observed that the value of $c_s^2$ depends significantly on which observable is used as a centrality classifier~\cite{Nijs:2023bzv}. 
It depends on whether one uses a charged multiplicity or a transverse energy, and whether the centrality detector overlaps with the analysis detector, in which $N_{ch}$ and $\langle p_T\rangle$ are measured. 

We first discuss, in Sec.~\ref{s:selfcorr}, the simple case where the centrality classifier is $N_{ch}$ itself. 
The case of a ``third-party'' centrality estimator will be discussed below in Sec.~\ref{s:thirdparty}.

\subsection{Self-correlations}
\label{s:selfcorr}

The simplest choice is to use $N_{ch}$ itself as a centrality classifier. 
The price to pay is an effect referred to as a ``self correlation'', which we describe below, and can be corrected.

%However, this induces ``self correlations'', arising from the fact that the same particles are used in the centrality determination and in the analysis itself.

%Here we would like to argue that the analysis can still be carried out if centrality detector and the analysis detector overlap. 

%But the effect of the self-correlation can easily be corrected, as explained in Ref.~\cite{Gardim:2019brr}. 
%We illustrate this correction on the simple case where the centrality is determined from the charged multiplicity $N_{ch}$, so that the centrality detector and the analysis detector are identical. 
In experiment, the fluctuations of $N_{ch}$ at $b=0$ around its average value $\langle N_{ch}\rangle$ are partly statistical, partly dynamical. 
More specifically, the variance $\sigma_{N_{ch}}^2$ contains a trivial contribution $\langle N_{ch}\rangle$ corresponding to Poisson fluctuations. 
The dynamical fluctuation $\sigma_{dyn}$ is obtained after subtracting this contribution:  $\sigma_{dyn}^2=\sigma_{N_{ch}}^2-\langle N_{ch}\rangle$.
Both $\sigma_{N_{ch}}$ and $\langle N_{ch}\rangle$ at $b=0$ can be accurately reconstructed from the distribution of $p(N_{ch})$ by simple Bayesian inference~\cite{Das:2017ned}, so that $\sigma_{dyn}$ can be inferred from data alone~\cite{Yousefnia:2021cup}. 

The increase of $\langle p_T\rangle$ is driven by dynamical fluctuations only.
Therefore, the derivative of $\langle p_T\rangle$ with respect to $N_{ch}$ is decreased by a factor $\sigma_{dyn}/\sigma_{N_{ch}}$ as the result of statistical fluctuations. 
As a consequence, the value of $c_s$ extracted from Eq.~(\ref{csdata}) is too low~\cite{Nijs:2023bzv}, and must be corrected accordingly: 
\begin{equation}
\label{cspoisson}
c_s^2(T_{\rm eff})=\left(1-\frac{\langle N_{ch}\rangle}{\sigma_{N_{ch}}^2}\right)^{-1/2}\frac{d\ln\langle p_T\rangle }{d\ln N_{ch}}. 
\end{equation} 
Let us evaluate the order of magnitude of this correction. 
In the central rapididy window, the relative dynamical variance $\sigma_{dyn}^2/\langle N_{ch}\rangle^2$ is approximately equal to $2\times 10^{-3}$ in Pb+Pb collisions at $\sqrt{s_{\rm NN}}=5.02$~TeV~\cite{Yousefnia:2021cup}.
Thus the correction factor in Eq.~(\ref{cspoisson}) is roughly $\sqrt{1+500/N_{ch}}$, which is not negligible if less than 2000 particles are seen~\cite{ALICE:2015juo}.
But it is an easy correction, which can be generalized to the case where there the centrality detector and the analysis detector partially overlap. 

For the sake of precision, let us mention that statistical fluctuations can differ from Poisson fluctuations due to short-range correlations~\cite{Bzdak:2012tp}, referred to as ``nonflow'' in the context of anisotropic flow studies~\cite{PHOBOS:2010ekr,STAR:2011ert}.
The variance associated with these short-range correlations is however a small relative correction to the variance of Poisson fluctuations, in the same way as nonflow correlations are small compared to self-correlations. 

\subsection{Kinematic cuts}
\label{s:kinematic}

The next difficulties are associated with the acceptance of the analysis detector. 
Ideally, one should measure the average transverse momentum of {\it all\/} particles within a rapidity window, as the idea is to trace the total energy and entropy of the fluid, according to Eq.~(\ref{effective}). 
Experimentally, however, one only measures the momenta of  {\it charged} particles. 
We have checked that $\langle p_T\rangle$ is unchanged in our calculations (within error bars) whether or not neutral particles are included. 
Therefore, no precision is lost by working with charged particles only. 

More importantly, experiments only detect particles whose transverse momentum exceeds some threshold (0.3~GeV/c in the case of CMS), because low-momentum particles are bent by the magnetic field too strongly and do not reach the inner detector.
The effect of such a low-$p_T$ cut on $\langle p_T\rangle$ was already pointed out by Nijs and van der Schee~\cite{Nijs:2023bzv}. 
The effect of an upper $p_T$ cut is also larger than one might naively expect: 
The reason is that in hydrodynamics, the increase of $\langle p_T\rangle$ resulting from an increase in $N_{ch}$ involves an intermediate agent, which is the fluid velocity. 
The increase in $N_{ch}$ implies a larger initial temperature, which in turn implies a larger fluid velocity at freeze-out because the hydrodynamic expansion lasts longer.
A fluctuation in the fluid velocity modifies the slope of the $p_T$ spectrum, and its effect is largest at high $p_T$~\cite{Gardim:2019iah}. 

Note that we do not yet know how the increase of $\langle p_T\rangle$ is distributed as a function of $p_T$. 
A closely-related question is how event-to-event fluctuations in the $p_T$ per particle are distributed in $p_T$. 
An observable has been devised for this purpose and dubbed $v_0(p_T)$ by Schenke, Shen and Teaney~\cite{Schenke:2020uqq}. 
It is still awaiting an experimental analysis. 

Application of Eq.~(\ref{csdata}) requires that $\langle p_T\rangle $ is evaluated without any cut. 
Therefore, it is essential to extrapolate the measured spectra before applying Eqs.~(\ref{ptteff}) and (\ref{csdata}). 
This extrapolation is carried out by CMS on the basis of a generic parameterizations of the $p_T$ spectrum~\cite{CMS:2024sgx}. 

The robustness of Eq.~(\ref{csdata}) against theoretical uncertainties is largely due to the fact that $N_{ch}$ and $\langle p_T\rangle$ are global observables~\cite{Noronha-Hostler:2015uye}. 
Spectra are not as well understood. 
Calculations underestimate the yield of low-$p_T$  pions~\cite{Grossi:2020ezz,Grossi:2021gqi,Guillen:2020nul}, and are in general less accurate for $p_T>2$~GeV/$c$~\cite{ALICE:2019hno}.
In these regions, model predictions are sensitive to the hadronization mechanism, therefore, some of the existing discrepancies can be attributed to our incomplete understanding of the hadronization phase. 
Uncertainties are much reduced upon integrating over phase space and summing over particle species.  

The place where hydrodynamics is expected to fail is the high $p_T$ range (larger than a few GeV/$c$). 
Experimental data on anisotropic flow~\cite{ALICE:2022zks} can only be explained by models which explicitly include a non-hydrodynamic component~\cite{Zhao:2021vmu}. 
This component consists of particles emitted before the fluid is formed~\cite{Kanakubo:2021qcw} or of jets propagating through the fluid. 
Even though this non-hydrodynamic component represents a very small fraction of the total, the resulting uncertainty on the speed of sound should eventually be estimated. 

\begin{figure}[h]
\begin{center}
\includegraphics[width=\linewidth]{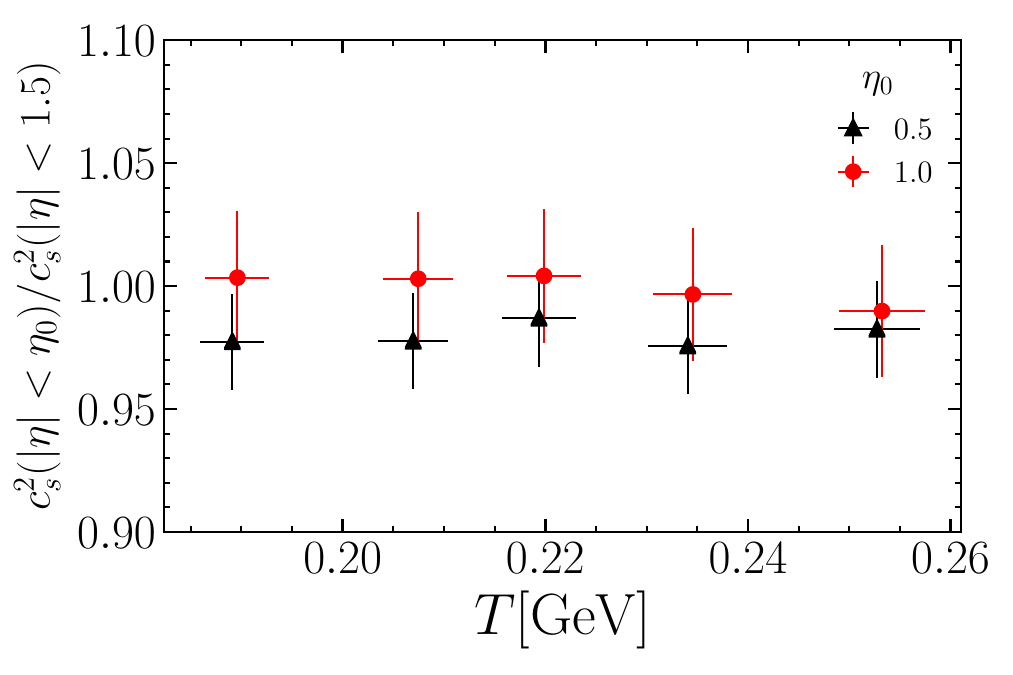} 
\end{center}
\caption{
Effect of the cut in pseudorapidity $\eta$. The speed of sound squared is evaluated using Eq.~(\ref{csdata}) with various cuts in pseudorapidity, and we plot the ratio of the resulting values to those shown in Fig.~\ref{fig:cs2}. Vertical error bars are statistical and horizontal error bars are as in Fig.~\ref{fig:cs2}.
}
\label{fig:etacut}
\end{figure} 

The second kinematic cut is the pseudorapidity selection. 
Ideally, for a boost-invariant fluid, one should evaluate $\langle p_T\rangle$ for all particles in a rapidity window. 
The pseudorapidity $\eta$ is identical to the rapidity $y$ for a particle moving close to the speed of light, but one has in general $|\eta|>|y|$, where the relative difference increases as $p_T/m$ decreases.
Thus, a narrower interval in $\eta$ misses some of the particles with lower $p_T$, which results in an increase of $\langle p_T\rangle$.
This is a modest effect, but we find that $\langle p_T\rangle/T_{\rm eff}$ increases by $\sim 1.5\%$ as the pseudorapidity window is narrowed from $|\eta|<1.5$ down to $|\eta|<0.5$.
This effect becomes milder as the temperature increases and particles velocities increase. 
The net effect is that the resulting estimate of the speed of sound using Eq.~(\ref{csdata}) is slightly reduced, as illustrated in Fig.~\ref{fig:etacut}. 
The effect is smaller than the error bar reported by CMS~\cite{CMS:2024sgx}, so that the cut $|\eta|<0.5$ implemented in the analysis does not hinder the precision of the result. 

\subsection{Local density fluctuations}
\label{s:localfluct}

Last but not least, the initial conditions of our hydrodynamic calculations are not realistic. 
We have assumed for simplicity that the density profile is smooth. 
Event-by-event fluctuations of the initial density profile~\cite{Aguiar:2001ac} play an essential role in the phenomenology of collective effects, in particular anisotropic flow~\cite{PHOBOS:2006dbo,Alver:2010gr}, and event-to-event transverse momentum fluctuations~\cite{Broniowski:2009fm,Bozek:2012fw}.

The question is how initial fluctuations affect our hypotheses, that $\langle p_T\rangle/T_{\rm eff}$ and $V_{\rm eff}$ are independent of the multiplicity. 
Hydrodynamic calculations with event-by-event fluctuations have shown that the correlation between $\langle p_T\rangle$ and $T_{\rm eff}$ is still remarkably strong in the presence of fluctuations, and that the value of $\langle p_T\rangle/T_{\rm eff}$ does not change significantly compared to smooth initial conditions~\cite{Gardim:2020sma}. 

On the other hand, the assumption that $V_{\rm eff}$ remains constant in the presence of fluctuations has been questioned by Nijs and van der Schee~\cite{Nijs:2023bzv}, who argue that it depends on how one models initial density fluctuations.  
They point out that high-multiplicity collisions might go along with a smaller transverse area, implying a smaller $V_{\rm eff}$.
This would imply that the effective entropy density increases faster than $N_{ch}$, so that Eq.~(\ref{csdata}) overestimates the speed of sound. 

This is definitely an interesting possibility that should be considered. 
There is at present little information on how fluctuations are distributed across the collision volume or, equivalently, about the two-point function of the initial energy density field~\cite{Blaizot:2014nia,Albacete:2018bbv}. 
The only quantitative information is about the rapidity dependence of fluctuations. 
Specifically, it has been shown that (relative) multiplicity fluctuations are larger at rapidities where the average multiplicity is itself larger,  both in Pb+Pb collisions~\cite{Yousefnia:2021cup} and in p+Pb collisions~\cite{Pepin:2022jsd} at zero impact parameter. 
%More specifically, in Pb+Pb collisions, the multiplicity density is smaller at large rapidities than around central rapidity, and also fluctuates less. 
%This results in the transverse energy $E_T$, measured at large rapidities, being a better centrality estimator than the charged multiplicity measured around central rapidity~\cite{ATLAS:2019peb}. 
%In p+Pb collisions, the multiplicity density is larger in the Pb-going direction than around central rapidity. 
%But the transverse energy $E_T$ measured in the Pb-going direction fluctuates more, in relative value, than the multiplicity around mid rapidity~\cite{ATLAS:2014qaj,Pepin:2022jsd}. 

As far as we know, there is no such knowledge about how fluctuations are distributed through the transverse plane. 
It could be that fluctuations are larger in the center where the density is largest~\cite{Giacalone:2019kgg}, which is the possibility mentioned by Nijs and Van der Schee. 
It could also be that they are smaller because the density of participants is so large that there is a saturation effect. 
This is one of the open questions in the field, with implications as for the phenomenology, in particular for anisotropic flow studies~\cite{Giacalone:2019kgg,Giannini:2022bkn}.
Since it is unclear whether the volume becomes larger or smaller in ultracentral collisions, the conservative assumption is that it is constant, as assumed in Ref.~\cite{Gardim:2019brr} and in the present calculation.
Experiment may actually shed light on this question. 
%One can compare the increase of $\langle p_T\rangle$ resulting from an increase in the collision energy to that resulting from an increase of multiplicity at fixed collision energy. 
%One can study and compare the increase in $\langle p_T\rangle$ in two distinct methods: by increasing the collision energy in 0-5$\%$ central collisions to eliminate initial condition fluctuations, or by studying ultracentral collisions at fixed collision energy, where the fluctuations increase the energy density. 
%As explained in Sec.~\ref{s:method}, both are equivalent in our setup with smooth initial conditions, but they are not obviously equivalent in the presence of fluctuations. 
Instead of studying ultracentral collisions at fixed collision energy, one can choose a fixed centrality range (say, 0-5\%), and vary the collision energy. 
Both $N_{ch}$ and $\langle p_T\rangle$ increase as a function of collision energy. 
It is likely that the effective volume remains essentially constant,\footnote{Note, however, that as one increases the collision energy, the nucleon-nucleon cross section increases, which result in a slight increase of the collision volume, which we neglect.} so that Eq.~(\ref{csdata}) applies. 
It is interesting to note that the value of $c_s^2$ inferred from the increase in collision energy from 2.76~TeV to 5.02~TeV per nucleon pair~\cite{Gardim:2019xjs} is compatible with that obtained by CMS~\cite{CMS:2024sgx} in ultracentral collisions, which seems to support the hypothesis that the effective volume is insensitive to local fluctuations. 
%It is interesting to note that the recent CMS analysis of ultracentral collisions~\cite{CMS:2024sgx} returns an increase of $\langle p_T\rangle$ which is precisely the same as that inferred from the increase in collision energy from 2.76~TeV to 5.02~TeV per nucleon pair~\cite{Gardim:2019xjs}, so that there is at present no hint that fluctuations are not distributed homogeneously.
 
Another consequence of local density fluctuations is that the relation between $N_{ch}$ and $\langle p_T\rangle$ is no longer one-to-one in a hydrodynamic simulation. 
For a fixed $N_{ch}$, there are dynamical fluctuations of $\langle p_T\rangle$~\cite{Broniowski:2009fm,Bozek:2012fw}. 
The standard deviation of $\langle p_T\rangle$ is of order $1\%$ in ultracentral collisions~\cite{ALICE:2014gvd,ATLAS:2022dov,ATLAS:2023xpw,ALICE:2023tej}, while that of $N_{ch}$ is of order $4\%$~\cite{Yousefnia:2021cup}. 
One can reasonably hope that Eq.~(\ref{csdata}) still holds after averaging over fluctuations, but how precisely should be assessed through a dedicated study, which we leave for future work. 

\subsection{Third-party centrality estimator}
\label{s:thirdparty}

%Finally, the local fluctuations studied in Sec.~\ref{s:localfluct} may affect the determination of the speed of sound, depending on the chosen centrality estimator. 

Most heavy-ion analyses are carried out using a centrality detector which does not overlap with the detectors where other quantities of interest are measured~\cite{ALICE:2013hur}. 
This is the case for the CMS analysis, where the centrality estimator is the transverse energy collected in two forward calorimeters~\cite{CMS:2024sgx}. 

The centrality estimator is, in general, a third observable, in addition to $N_{ch}$ and $\langle p_T\rangle$. 
The local fluctuations mentioned in Sec.~\ref{s:localfluct} induce dynamical fluctuations of the centrality estimator, whose correlation with $N_{ch}$ and $\langle p_T\rangle$ are in general not trivial. 
If the centrality estimator is itself a charged multiplicity, one may hope that it is correlated to $N_{ch}$ rather than $\langle p_T\rangle$, so that Eq.~(\ref{csdata}) applies. 
When the centrality estimator is a transverse energy, on the other hand, it is natural to expect that it is positively correlated with $\langle p_t\rangle$, so that Eq.~(\ref{csdata}) would overestimate $c_s$~\cite{Nijs:2023bzv}. 
It should be possible to systematically assess the effect of this bias, which we also leave up for future work.

%It is important to compare different centrality classifiers in order to assess the robustness of the result. 
%The choice made by CMS~\cite{CMS:2024sgx} is to determine the centrality in a specific detector, which we refer to as the centrality detector, and measure  $\langle p_T\rangle$ versus $dN_{ch}/d\eta$ in a separate detector, which we refer to as the analysis detector. 

\section{Conclusion}

The conclusion of our work is that the determination of the speed of sound from experimental data suggested in Ref.~\cite{Gardim:2019brr} and implemented in Ref.~\cite{CMS:2024sgx} seems robust and, perhaps surprisingly, precise, within a hydrodynamic description with smooth initial conditions. 

The only serious caveat comes from the locality of  initial fluctuations, which has a number of effects: 
It potentially induces a bias when the centrality is estimated using a transverse energy, as discussed in Sec.~\ref{s:thirdparty}. 
Local fluctuations may also result in an effective volume which depends on the multiplicity~\cite{Nijs:2023bzv}. 
Both effects deserve further investigations, which will shed new light on initial fluctuations. 

In addition, there is a limit on the precision of the measurement of the speed of sound from the hydrodynamic description itself. 
In particular, it will be important to assess quantitatively the error stemming from the non-hydrodynamic production at high $p_T$.  

\begin{acknowledgments}
We thank Cesar Bernardes, Alexander Kalweit, Wei Li, Jurgen Schukraft, Govert Nijs, Wilke van der Schee, Marco van Leeuwen, Krishna Rajagopal, Anthony Timmins and Urs Wiedemann for discussions. 
FGG is supported by CNPq (Conselho Nacional de Desenvolvimento Cientifico) through 306762/2021-8, INCT-FNA grant 312932/2018-9 and 409029/2021-1 and Fulbright Program.The authors acknowledge the National Laboratory for Scientific Computing (LNCC/MCTI, Brazil), through the ambassador program (UFGD), subproject FCNAE, for providing HPC resources of the SDumont supercomputer, which have contributed to the research results reported within this study. URL: \url{http://sdumont.lncc.br}.
\end{acknowledgments}

\end{document}